\documentclass{aa}
\usepackage{graphicx,times}
\usepackage{amsmath,amsfonts}

\def\n{{\bf n}}
\def\x{{\bf x}}
\def\v{{\bf v}}

\def\T{{\cal T}}
\def\D{{\cal D}}
\def\I{{\cal I}}
\def\S{{\cal S}}
\def\A{{\cal A}}

\def\s{\sigma}

\begin{document}

   \title{Radiative Transfer with Finite Elements}

   \subtitle{II. Ly$\alpha$ Line Transfer in Moving Media}

   \author{E. Meink\"ohn\inst{1,2}
          \and
          S. Richling\inst{1}
          }

   \offprints{S. Richling}

   \institute{Institut f\"ur Theoretische Astrophysik,
              Universit\"at Heidelberg,
              Tiergartenstr. 15,
              D-69121 Heidelberg
         \and
             Institut f\"ur Angewandte Mathematik,
             Universit\"at Heidelberg,
             Im Neuenheimer Feld 294,
             D-69120 Heidelberg
             }

   \date{Received Month Day, 2002; accepted Month Day, 2002}

   \abstract{A finite element method for solving the resonance line
    transfer problem in moving media is presented. The algorithm works
    in three spatial dimensions on unstructured grids which are
    adaptively refined by means of an a posteriori error
    indicator. Frequency discretization is implemented via a
    first-order Euler scheme. We discuss the resulting matrix
    structure for coherent isotropic scattering and complete
    redistribution. The solution is performed using an iterative
    procedure, where monochromatic radiative transfer problems are
    successively solved. The present implementation is applicable for
    arbitrary model configurations with an optical depth up to
    $10^{3-4}$.  Results of Ly$\alpha$ line transfer calculations for
    a spherically symmetric model, a disk-like configuration, and a
    halo containing three source regions are discussed. We find the
    characteristic double-peaked Ly$\alpha$ line profile for all
    models with an optical depth $\ga 1$. In general, the blue peak of
    the profile is enhanced for models with infall motion and the red
    peak for models with outflow motion. Both velocity fields produce
    a triangular shape in the two-dimensional Ly$\alpha$ spectra,
    whereas rotation creates a shear pattern. Frequency-resolved
    Ly$\alpha$ images may help to find the number and position of
    multiple Ly$\alpha$ sources located in a single halo. A
    qualitative comparison with observations of extended Ly$\alpha$
    halos associated with high redshift galaxies shows that even
    models with lower hydrogen column densities than required from
    profile fitting yield results which reproduce many features in the
    observed line profiles and two-dimensional
    spectra.  }

    \maketitle

\section{Introduction}

Hydrogen Ly$\alpha$ as a prominent emission line of high redshift
galaxies is important for the understanding of galaxy formation and
evolution in the early universe. Recent narrow-band imaging and
spectroscopic surveys used the Ly$\alpha$ line to identify galaxies at
very high redshift (e.g. Hu et al. \cite{hu:etal98}; Kudritzki et
al. \cite{kudritzki:etal2000}; Rhoads et al. \cite{rhoads:etal2000};
Fynbo et al. \cite{fynbo:etal2000}, \cite{fynbo:etal2001}).  

But beside from being a good redshift indicator, Ly$\alpha$ emission
also bears information on the distribution and kinematics of the
interstellar gas as well as the nature of the energy
source. Ly$\alpha$ observations of high redshift radio galaxies
(e.g. Hippelein \& Meisenheimer \cite{hippelein:meisenheimer93}; van
Ojik et al. \cite{vanojik:etal96}, \cite{vanojik:etal97};
Villar-Mart{\'{\i}}n et al. \cite{villar-martin:etal99}) reveal
extended Ly$\alpha$ halos with sizes $>100$~kpc which are aligned with
the radio jet. The line profiles are often double-peaked and the
two-dimensional spectra point to complex kinematics involving
velocities $>1000$~km$\;\mbox{s}^{-1}$.

The interpretation of Ly$\alpha$ observations is difficult, because
high-redshift radio galaxies tend to be in the center of proto
clusters, where the radio jet interacts with a clumpy environment
influenced by merging processes (e.g. Bicknell et
al. \cite{bicknell:etal2000}; Kurk et
al. \cite{kurk:etal2001}). Actually, the three-dimensional structure
of the objects and the fact that Ly$\alpha$ is a resonance line
require detailed radiative transfer modeling.

The transfer of resonance line photons is profoundly determined by
scattering in space and frequency. Analytical (Neufeld
\cite{neufeld90}) as well as early numerical methods (Adams
\cite{adams72}; Hummer \& Kunasz \cite{hummer:kunasz80}) were
restricted to one-dimensional, static media. Only recently, codes
based on the Monte Carlo method were developed which are capable to
investigate the more general case of a multi-dimensional medium (Ahn
et al. \cite{ahn:etal2001}, \cite{ahn:etal2002}).

In this paper, we introduce a finite element method for the solution
of the resonance line transfer problem in three-dimensional, moving
media and present the results of some simple, slightly optically thick
model configurations. The basic, monochromatic code was originally
developed by Kanschat (\cite{kanschat96}) and is described in Richling
et al. (\cite{richling:etal2001}, hereafter Paper~I). The
three-dimensional method is particularly useful for scattering
dominated problems in inhomogeneous media. Steep gradients are
resolved by means of an adaptively refined grid. Here, we only specify
the extension from the monochromatic to the polychromatic problem
including the implementation of the Doppler-effect and complete
redistribution.

In Sect.~2, we review the equations for resonance line transfer in
moving media. In Sect.~3, we describe some details regarding frequency
discretization and the form of the resulting matrices for coherent
scattering and complete redistribution and give a short outline of the
complete finite element algorithm. Then, the results of a spherically
symmetric model (Sect.~4), a disk-like configuration (Sect.~5) and a
model with three separate source regions (Sect.~6) are presented. A
summary is given in Sect.~7.

\section{Line Transfer in Moving Media}
We calculate the frequency-dependent radiation field in moving media
by solving the non-relativistic radiative transfer equation in the
comoving frame for a three-dimensional domain $\Omega$ which in
operator form simply reads
\begin{equation}\label{fqrt}
\bigl(\T+\D+\S+\chi(\x,\nu)\bigr)\I(\x,\n,\nu)=f(\x,\nu).
\end{equation}
The transfer operator $\T$, the ``Doppler'' operator $\D$, and the
scattering operator $\S$ are defined by
\begin{eqnarray}
\lefteqn{\T\I(\x,\n,\nu)=\n\cdot\nabla_x \I(\x,\n,\nu),}\nonumber\\
\lefteqn{\D\I(\x,\n,\nu)=
       w(\x,\n)\,\nu\frac{\partial}{\partial\nu}\I(\x,\n,\nu),}\nonumber\\
\lefteqn{\S\I(\x,\n,\nu)=
       -\frac{\sigma(\x)}{4\pi}
       \int_{0}^{\infty} \int_{S^2} R(\hat{\n},\hat{\nu};\n,\nu)
        \I(\x,\hat{\n},\hat{\nu})\,d\hat{\omega}\,d\hat{\nu}.}\nonumber
\end{eqnarray}
Considering three dimensions in space, the relativistic invariant
specific intensity $\I$ is six-dimensional and depends on the space
variable $\x$, the direction $\n$ (pointing to the solid angle
$d\omega$ of the unit sphere $S^2$), and the frequency $\nu$.

The extinction coefficient
$\chi(\x,\nu)=\kappa(\x,\nu)+\sigma(\x,\nu)$ is the sum of the
absorption coefficient $\kappa(\x,\nu)=\kappa(\x)\phi(\nu)$ and the
scattering coefficient $\sigma(\x,\nu)=\sigma(\x)\phi(\nu)$. The
frequency-dependence is given by a
normalized profile function $\phi$.  Usually, we adopt a Doppler
profile
\begin{equation}\label{doppler}
\phi(\nu)=\frac{1}{\sqrt{\pi}\,\Delta\nu_{D}}
\exp{\left[-\left(\frac{\nu-\nu_0}{\Delta\nu_{D}}\right)^2\right]},
\end{equation}
where $\nu_0$ is the frequency of the line center. The Doppler width
$\Delta\nu_{D}$ and the Doppler velocity $v_D$ are determined by a
thermal velocity $v_{\rm therm}$ as well as a turbulent velocity
$v_{\rm turb}$
\begin{equation}
\Delta\nu_{D}=\frac{\nu_0}{c} v_D=
\frac{\nu_0}{c} \sqrt{v_{\rm therm}^2+v_{\rm turb}^2}.
\end{equation}
$c$ is the speed of light. For situations, where $v_{\rm turb}\gg
v_{\rm therm}$, the Doppler core is very broad and would dominate the
Lorentzian wings of a Voigt profile. Then, the Doppler profile is a
reasonable description of a line profile. Throughout this paper we
use $v_{D}=10^{-3}c\sim300\;\mbox{km}\,\mbox{s}^{-1}$ (see also
Table~\ref{parameters_table}).

For the source term
\begin{equation}
f(\x,\nu)=\kappa(\x,\nu)B(T(\x),\nu)+\epsilon(\x,\nu),
\end{equation}
we can consider thermal radiation and non-thermal radiation.  In the
case of thermal radiation, $f$ is calculated from a temperature
distribution $T(\x)$, where $B(T,\nu)$ is the Planck function.

The ``Doppler'' operator $\D$ is responsible for the Doppler shift of
the photons. A derivation of the operator for non-relativistic
velocities ($v/c < 0.1$) can be found in Wehrse et
al. (\cite{wehrse:etal2000}). In contrast to the full relativistic
transfer equation (cf.  Mihalas \& Weibel-Mihalas
\cite{mihalas:weibel-mihalas84}), we neglect any terms responsible for
aberration or advection effects. The function
\begin{equation}\label{w_fkt}
w(\x,\n)=-\n\cdot\nabla_x \left(\n\cdot \frac{\v(\x)}{c} \right) 
\end{equation}
is the gradient of the velocity field $\v(\x)$ in direction $\n$. Note
that the sign of $w$ may change depending on the complexity of the
velocity field $\v$.

The scattering operator $\S$ depends on the general redistribution
function $R(\hat{\n},\hat{\nu};\n,\nu)$ giving the probability that
a photon $(\hat{\n},\hat{\nu})$ is absorbed and a photon $(\n,\nu)$ is
emitted. In the following, we assume isotropic scattering
\begin{equation}
\S \I = -\frac{\s(\x)}{4\pi}
           \int\limits_{0}^{\infty} R(\hat{\nu},\nu)
           \int_{S^2} \I(\x,\hat{\n},\hat{\nu})\,d\hat{\omega}\,d\hat{\nu},
\end{equation}
where $R(\hat{\nu},\nu)$ is the angle-averaged redistribution function
\begin{equation}\label{aar}
R(\hat{\nu},\nu)=\frac{1}{(4\pi)^2}\int_{S^2}\int_{S^2}
R(\x,\hat{\n},\hat{\nu};\n,\nu)\,d\hat{\omega}\,d\omega.
\end{equation}
The function defined by (\ref{aar}) is normalized such that
\begin{equation}
\int\limits_0^\infty \int\limits_0^\infty
R(\hat{\nu},\nu)\,d\hat{\nu}\,d\nu = 1.
\end{equation}
For the calculations in Sect.~3, we consider two limiting cases:
strict coherence and complete redistribution in the comoving frame. In
the former case, we have
\begin{equation}
R(\hat{\nu},\nu) = \phi(\hat{\nu}) \delta (\nu-\hat{\nu})
\end{equation}
and in the latter
\begin{equation}
R(\hat{\nu},\nu) = \phi(\hat{\nu}) \phi(\nu).
\end{equation}
Thus, for coherent isotropic scattering, the scattering term
simplifies to
\begin{equation}
\S^{\rm coh}\I(\x,\n,\nu)=-\frac{\sigma(\x,\nu)}{4\pi}
        \int_{S^2}\I(\x,\hat{\n},\nu)\,d\hat{\omega}   
\end{equation}
In the case of complete redistribution, the photons are scattered
isotropically in angle, but are randomly redistributed over the line
profile. Then, the scattering term reads
\begin{equation}\label{scat_crd}
\S^{\rm crd}\I(\x,\n,\nu) = 
-\frac{\sigma(\x,\nu)}{4\pi} \int_0^{\infty}\phi(\hat{\nu})
\int_{S^2} \I(\x,\hat{\n},\hat{\nu})\,d\hat{\omega}\,
d\hat{\nu}. 
\end{equation}

\section{The Finite Element Method}

\subsection{Boundary Conditions}
For the modeling of prominent resonance lines, in particular
Ly$\alpha$, we restrict the frequency discretization to a finite
interval $\Lambda:=[\nu_0,\nu_{N+1}]$, where $\nu_0$ and $\nu_{N+1}$
are located far out of the line center in the continuum. The function
$w(\x,\n)$ from Eq.\ (\ref{w_fkt}) defines the Doppler shift of the
spectrum towards smaller ($w<0$) or larger ($w>0$) frequencies in the
comoving frame. Therefore, boundary conditions of the form
\begin{equation*}
\I(\x,\n,\nu)=\I_{\rm cont}(\x,\n,\nu)
\end{equation*}
are necessary on the upper and lower frequency interval boundary of
the whole computational domain
\begin{equation*}
\Sigma = \Omega \times \S^2 \times \{\nu_0,\nu_{N+1}\}.
\end{equation*}
Furthermore, to be able to solve Eq.\ (\ref{fqrt}), boundary
conditions of the form
\begin{equation*}
\I(\x,\n,\nu)=\I_{\rm in}(\x,\n,\nu) 
\end{equation*}
must be imposed on the ``inflow boundary'' 
\begin{equation*}
\Gamma^- \times \Lambda = \bigl\{ (\x,\n,\nu) \in \Gamma | \mbox{\,} \n_\Gamma 
\cdot \n < 0 \bigr\},
\end{equation*}
where $\n_\Gamma$ is the unit vector perpendicular to the boundary
surface $\Gamma$ of the spatial domain $\Omega$. The sign of the
product $\n_\Gamma \cdot \n$ describes the ``flow direction'' of the
photons across the boundary. If we neglect any continuum emission
($\I_{\rm cont}=0$) and assume that there are no light sources outside
the modeled domain as in the case of a non-illuminated atmosphere
($\I_{\rm in}=0$), the two boundary conditions for the solution of the
transfer equation (\ref{fqrt}) in moving media are
\begin{align}
\I(\x,\n,\nu) & = 0 \mbox{\qquad on }\; \Sigma , \\
\I(\x,\n,\nu) & = 0 \mbox{\qquad on }\; \Gamma^- \times \Lambda .
\end{align}

\subsection{Discretization for Coherent Scattering}
In order to solve the radiative transfer equation (\ref{fqrt}), we
first carry out the frequency discretization including coherent
scattering. With the abbreviation
\begin{equation}
\A^{\rm coh}=\T+\S^{\rm coh}+\chi(\x,\nu)
\end{equation}
Eq.~(\ref{fqrt})  can be written as
\begin{equation}\label{coh_rte}
\A^{\rm coh}\I(\x,\n,\nu)+w(\x,\n)\nu\frac{\partial}{\partial \nu}\I(\x,\n,\nu)
=f(\x,\nu).
\end{equation}
In Paper~I, we described a Galerkin discretization for the
monochromatic radiative transfer equation. Considering memory and CPU
requirements this approach is by far too ``expensive'' for the
frequency-dependent transfer problem. Instead, we use an implicit
Euler method for $N$ equidistantly distributed frequency points $\nu_i
\in \bigl\{ \nu_1,\nu_2,...,\nu_N \bigr\} \subset \Lambda$. Since the
function $w(\x,\n)$ may change its sign, this simple difference scheme
for the Doppler term reads
\begin{align}
w(\x,\n)\nu\frac{\partial \I}{\partial \nu} & \longrightarrow 
w\nu_i \frac{\I_i - \I_{i-1}}{\Delta \nu} \mbox{\qquad} (w_i>0) \\
\intertext{and}
w(\x,\n)\nu\frac{\partial \I}{\partial \nu} & \longrightarrow 
w\nu_i \frac{\I_{i+1} - \I_i}{\Delta \nu} \mbox{\qquad} (w_i<0), 
\end{align}
where $\Delta \nu$ is the constant frequency step size. All quantities
referring to the discrete frequency point $\nu_i$ are denoted by an
index ``$i$''. Employing the Euler method, we get a semi-discrete
representation of Eq.\ (\ref{coh_rte})
\begin{equation}\label{semi_coh}
\Bigl(\A_i^{\rm coh}+\frac{|w|\nu_i}{\Delta \nu}\Bigr)\I_i=
      f_i+\frac{|w|\nu_i}{\Delta \nu}
      \left\{ \begin{array}{r@{\quad}l}
       \I_{i-1} & (w > 0) \\ \I_{i+1} & (w < 0) 
       \end{array} \right. .
\end{equation}
The additional term on the left side of Eq.\ (\ref{semi_coh}) can be
interpreted as an artificial opacity, which is advantageous for the
solution of the resulting linear system of equations. The additional
term on the right side of Eq.~(\ref{semi_coh}) is included as an
artificial source term. Equation~(\ref{semi_coh}) can be written in a
compact operator form
\begin{equation}\label{cohrt}
\tilde{\A}_i^{\rm coh}\I_i=\tilde{f_i}.
\end{equation}
Given a discretization with $L$ degrees of freedom in $\Omega$, $M$
directions on the unit sphere $S^2$ and $N$ frequency points, the
overall discrete system has the matrix form
\begin{equation}\label{block_coh}
{\bf A}^{\rm coh} {\bf u} = {\bf f} ,
\end{equation}
with the vector ${\bf u}$ containing the discrete intensities and the
vector ${\bf f}$ the values of the source term for all frequency
points $\nu_i$. Both vectors are of length $(L \cdot M \cdot N)$ and
${\bf A}^{\rm coh}$ is a $(L \cdot M \cdot N)\times(L \cdot M \cdot
N)$ matrix. The fact that the function $w(\x,\n)$ may change its sign,
results in a block-tridiagonal structure of ${\bf A}^{\rm coh}$ and we
get
\begin{multline*}
\begin{array}{cccc}
  \left( \begin{array}{ccccc} 
  {\tilde {\bf A}}_1^{\rm coh} & {\bf R}_1 & {\bf 0} & \ldots & {\bf 0} \\ 
  {\bf B}_2 & {\tilde {\bf A}}_2^{\rm coh} & {\bf R}_2 & \ddots & \vdots \\ 
  {\bf 0} & \ddots & \ddots & \ddots & {\bf 0} \\
  \vdots & \ddots & \ddots & \ddots & {\bf R}_{N-1} \\ 
  {\bf 0} & \ldots & {\bf 0} & {\bf B}_N & {\tilde {\bf A}}_N^{\rm coh} 
  \end{array} \right) & \left( \begin{array}{c} {\bf u}_1 \\ {\bf u}_2 \\ \vdots
  \\ \vdots \\ {\bf u}_N \end{array} \right) & = &
\left(
  \begin{array}{c}
        {\bf f}_1 \\
        {\bf f}_2 \\
	\vdots \\
        \vdots \\
	{\bf f}_N
	\end{array}
	\right) .
\end{array} 
\end{multline*}
According to the sign of $w(\x,\n)$ the block matrices ${\bf R}_i$ and
${\bf B}_i$ hold entries of $w(\x,\n)\nu_i/\Delta\nu$ causing a
redshift and blueshift of the photons in the medium, respectively.
Requiring a reasonable resolution, the resulting linear system of
equations of the total system is too large to be solved
directly. Hence, we are treating $N$ ``monochromatic'' radiative
transfer problems
\begin{gather}\label{block_coh_inu}
{\tilde {\bf A}}_i^{\rm coh} {\bf u}_i = \tilde{{\bf f}}_i ,
\end{gather}
with a slightly modified right hand side
\begin{gather}
\tilde{{\bf f}}_i = {\bf f}_i + {\bf R}_i {\bf u}_{i+1} + 
{\bf B}_i {\bf u}_{i-1} .
\end{gather}
Note that using simple velocity fields (e.g.\ infall or outflow) the
sign of $w(\x,\n)$ is fixed and the matrix ${\bf A}^{\rm coh}$
has a block-bidiagonal structure. This is favorable for our solution
strategy, since we only need to solve Eq.~(\ref{block_coh_inu}) once
for each frequency.

\subsection{Discretization for Complete Redistribution}
Considering complete redistribution, we use an implicit Euler scheme to
discretize the Doppler term as described above and a simple quadrature
method for the frequency integral in the scattering operator $\S^{\rm
crd}$ in Eq.~(\ref{scat_crd}). Starting from $N$ equidistantly
distributed frequency points $\nu_i \in \bigl\{ \nu_1,\nu_2,...,\nu_N
\bigr\} \subset \Lambda$ and $N$ weights $q_1,q_2,...,q_N$, we define a
quadrature method
\begin{equation}
Q(\nu_i) := \sum_{j=1}^N q_j \xi(\nu_j) 
\end{equation}
for integrals $\int_{\Lambda} \xi(\nu')d\nu'$. In the case of complete
redistribution the kernel is
\begin{equation}
\xi(\nu_j) = \frac{\phi(\nu_j)}{4\pi}\int_{S^2}\I(\x,\hat{\n},\nu_j)
d\hat{\omega} .
\end{equation}
Separating the terms with the unknown intensities $\I_i$ from the
known quantities $\I_j$ the discretized scattering integral
(\ref{scat_crd}) reads
\begin{equation}
\frac{\sigma_i}{4\pi} \phi_i q_i \int_{S^2} \I_id\hat{\omega} + 
\frac{\sigma_i}{4\pi}\sum_{j\not=i}^N  \phi_j q_j \int_{S^2}
\I(\x,\hat{\n},\nu_j)d\hat{\omega} .
\end{equation}
Using this quadrature scheme and employing the Euler method for the
frequency derivatives we get a semi-discrete formulation of the
transfer problem for each frequency point including complete
redistribution
\begin{multline}\label{semi_crd}
\Bigl(\A^{\rm crd}_i+ \frac{|w|\nu_i}{\Delta\nu}\Bigr)\I_i
 \\
\mbox{\hfill} = \tilde{f}_i+\frac{\sigma_i}{4\pi}\sum_{j\not=i}^N
\phi_jq_j \int_{S^2}\I(\x,\hat{\n},\nu_j)d\hat{\omega} ,
\end{multline}
where
\begin{gather}
\A^{\rm crd}_i = \T+\chi_i+\phi_i q_i \S^{\rm coh} .
\end{gather}
The additional terms on the right hand side of Eq.~(\ref{semi_crd})
must be interpreted as artificial source terms. Eq.~(\ref{semi_crd})
can also be written in a compact operator form
\begin{equation}\label{crdrt}
\tilde{\A}_i^{\rm crd}\I_i=\hat{f_i}.
\end{equation}
The total discrete system has the matrix form (cf.\
Eq.~(\ref{block_coh}))
\begin{gather}
{\bf A}^{\rm crd} {\bf u} = {\bf f} .
\end{gather}
Unfortunately, the global frequency coupling via the scattering
integral (\ref{scat_crd}) results in a full block matrix and we
get
\begin{multline*}
\begin{array}{ccc}
{\bf A}^{\rm crd} & = &
  \left( \begin{array}{ccccc} 
  {\tilde{\bf A}}_1^{\rm crd}&{\bf R}_1+{\bf Q}_2&{\bf Q}_3&\ldots&{\bf Q}_N \\ 
  {\bf B}_2+{\bf Q}_1&{\tilde{\bf A}}_2^{\rm crd}&{\bf R}_2+{\bf Q}_3&\ddots&\vdots \\ 
  {\bf Q}_1 & \ddots & \ddots & \ddots & \vdots\\
  \vdots & \ddots & \ddots & \ddots & \vdots\\ 
  {\bf Q}_1&\ldots& \ldots& \ldots&{\tilde {\bf A}}_N^{\rm crd} 
  \end{array} \right) .
\end{array} 
\end{multline*}
${\bf B}_i$ and ${\bf R}_i$ again contain the contribution of the
Doppler factor $w(\x,\n)\nu_i/\Delta\nu$, whereas ${\bf Q}_j$ holds
the terms from the quadrature scheme.  As we already explained in the
case of coherent scattering, we do not solve the total system, but $N$
``monochromatic'' radiative transfer problems
\begin{gather}
{\tilde {\bf A}}_i^{\rm crd} {\bf u}_i =  \hat{{\bf f}}_i ,
\end{gather}
with a modified right hand side
\begin{gather}\label{block_coh_inu2}
\hat{{\bf f}}_i = {\bf f}_i + {\bf R}_i {\bf u}_{i+1} + 
{\bf B}_i {\bf u}_{i-1} + \sum_{j\not=i} {\bf Q}_j {\bf u}_j .
\end{gather}

\subsection{Full Solution Algorithm}
Eq.\ (\ref{cohrt}) and (\ref{crdrt}) are of the same form as the
monochromatic radiative transfer equation, $\A\I=f$, for which we
proposed a solution method based on a finite element technique in
Paper~I. The finite element method employs unstructured grids which
are adaptively refined by means of an a posteriori error estimation
strategy. Now, we apply this method to Eq.~(\ref{cohrt}) or
Eq.~(\ref{crdrt}). In brief, the full solution algorithm reads:
\begin{enumerate}
\item Start with $\I=0$ for all frequencies.
\item Solve Eq.~(\ref{cohrt}) or Eq.~(\ref{crdrt}) for $i=1,..,N$.
\item Repeat step 2 until convergence is reached.
\item Refine grid and repeat step 2 and 3.
\end{enumerate}
We start with a relatively coarse grid, where only the most important
structures are pre-refined, and assure that the frequency interval
$[\nu_1,\nu_N]$ is wide enough to cover the total line profile.  Then,
we solve Eq.~(\ref{cohrt}) or Eq.~(\ref{crdrt}) for each frequency
several times depending on the choice of the redistribution function
and the velocity field. During this fix point iteration, we monitor
the changes of the resulting line profile in a particular direction
$\n_{\rm out}$. Not until the line profile remains unchanged, we go
over to step 4 and refine the spatial grid. Again, we apply the fixed
fraction grid refinement strategy: The cells are ordered according to
the size of the local refinement indicator
$\eta_K=max(\eta_K(\nu_i))|_{\nu_i}$ and a fixed portion of the cells
with largest $\eta_K$ is refined. $\eta_K(\nu_i)$ is an indicator for
the error of the solution in cell $K$ at frequency $\nu_i$.

In Paper~I (Sect.~2.3), we introduced various possibilities to
determine $\eta_K$. If we use the total flux leaving the computational
domain in direction $\n_{\rm out}$ as the quantity whose error
functional is used to calculate $\eta_K$ (dual method), our
convergence criterion described above is perfectly reasonable. We
found that this criterion is also useful for the much simpler method
(L$^2$ method), where $\eta_K$ is determined from the global error
functional. In this case, the rate of convergence of the line profile
in the directions $\n \ne \n_{\rm out}$ and for the line profile in
direction $\n_{\rm out}$ are very similar. We are aware that this
result may not be valid for more complex model configuration.

\section{Model Configurations}
\subsection{Halo and Sources}
We investigate spherically symmetric and elliptical distributions of
the extinction coefficient $\chi(\x)=\chi(x,y,z)$ of the form
\begin{equation}
\chi(\x)=\left\{ \begin{array}{ll}
                 \chi_0/(1+\alpha r_c^2) & \quad\mbox{for}\; r \le r_c\\
                 \chi_0/(1+\alpha r^2) & \quad\mbox{for}\; r_c < r \le r_h\\
                 \chi_0/(1+\alpha r_h^2)/10^3 & \quad\mbox{for}\; r > r_h
                \end{array} \right. ,
\end{equation}
where $r^2=(x/a)^2+(y/b)^2+(z/c)^2$. Note, that the value of the
extinction coefficient is constant in the center within core radius
$r_c$ and outside halo radius $r_h$. If not stated otherwise, $\chi_0$
is determined from the line center optical depth
\begin{equation}
\tau=\int_{r_c}^{r_h}\chi(r)\phi(\nu_0)\;\n_{\rm thick}\,d\x
\end{equation}
between $r_c$ and $r_h$ along the most optically thick direction
$\n_{\rm thick}$. In total, the spatial distribution of $\chi$ is
determined by six parameters: the length of the half-axes $a$, $b$ and
$c$, the radii $r_c$ and $r_h$, the dimensionless parameter $\alpha$,
and the optical depth $\tau$. For $r_c$, $r_h$ and $\alpha$ we use the
values given in Table~\ref{parameters_table}.

Since we are predominantly interested in the transfer of radiation in
resonance lines like Ly$\alpha$, we assume $\sigma(\x)=\chi(\x)$ and
$\kappa(\x)=0$ for all calculation presented here. This restricts us
to the use of purely non-thermal source functions. In particular, we
consider one or several spatially confined source regions with radius
$r_s$ each centered at a position $\x_i$:
\begin{equation}
f(\x,\nu)=\left\{ \begin{array}{lr}
      \phi(\nu) & \quad\mbox{for}\quad |\x-\x_i| \le r_s\\
              0 & \quad\mbox{for}\quad |\x-\x_i| > r_s
             \end{array} \right. .
\end{equation}
The function $\phi(\nu)$ is the Doppler profile defined in
Eq.~(\ref{doppler}).

\subsection{Velocity Fields}

In general, the finite element code is able to consider arbitrary
velocity fields. For velocity fields defined on a discrete grid, for
example for velocity fields resulting from hydrodynamical simulations,
the velocity gradient in direction $\n$ must be obtained
numerically. Here, we use two simple velocity fields and calculate the
function $w$ analytically.

The first velocity field describes a spherically symmetric
inflow ($v_0<0$) or outflow ($v_0>0$) and is of the form
\begin{equation}\label{vio_eq}
\v_{\rm io}=v_0 \left(\frac{r_0}{r}\right)^l \frac{\x}{r},
\end{equation}
where $r=|\x|$ and $v_0$ the  scalar velocity at
radius $r_0$. The corresponding $w$ function is
\begin{equation}
w(\x,\n)=v_0 \left(\frac{r_0}{r}\right)^l
               \left(\frac{1}{r}-(l+1) \frac{|\n\x|}{r^3} \right).
\end{equation}

For the second velocity field, we assume rotation around
the $z$-axis
\begin{equation}
\v_{\rm rot}=v_0 \left(\frac{R_0}{R}\right)^l R^{-1}
\left(\begin{array}{rrr}y\\-x\\0\end{array}\right),
\end{equation}
where $R^2=x^2+y^2$ is the distance from the rotational
axis and $v_0$ the scalar velocity at distance $R_0$.
If $\n=(n_x,n_y,n_z)$, the $w$ function reads
\begin{equation}
w=v_0 \left(\frac{R_0}{R}\right)^l (l+1)
     \left(\frac{xy (n_y^2-n_x^2) + n_x n_y (x^2-y^2)}{R^3} \right).
\end{equation}
In the special case of a Keplerian velocity field ($l=0.5$), we obtain
\begin{equation}
w=v_0 \frac{3}{2} R_0^{0.5} R^{-3.5}
 \left[x y (n_y^2-n_x^2) + n_x n_y (x^2-y^2)\right]
\end{equation}
(see also Baschek \& Wehrse \cite{baschek:wehrse99}).  Again, the
values used for $v_0$, $r_0$ and $R_0$ are given in
Table~\ref{parameters_table}.

\subsection{Normalization}\label{normalization}
Since we use a normalized form of the equation of transfer, where the
computational domain is the unit cube, the results are valid for model
configurations with different length scales and for different resonance
lines. When $2x_{\max}$ is the size of the unit cube, coordinates in
physical units are simply obtained by the transformation $\x \rightarrow
\x\cdot x_{\rm max}$.

A particular line and the corresponding frequency scale must be
defined when transforming to the observers frame. From the solution
$\I(\x,\n,\nu)$ in the comoving frame, we obtain the solution
$\tilde{\I}(\x,\n,\tilde{\nu})$ in the observers frame by performing
the transformation
\begin{equation}
\tilde{\I}(x,\n,\tilde{\nu})=\I(\x,\n,\tilde{\nu})
    \left(\frac{\tilde{\nu}}{\nu}\right)^3,
\end{equation}
where
\begin{equation}
\tilde{\nu}=\nu\left(1+\n\cdot \frac{\v(\x)}{c}\right).
\end{equation}

The relation between optical depth $\tau$ and column density of the
scattering media depends on the central line absorption cross section
$\chi(\nu_0)$ and therefore on the Doppler velocity $v_D$. For the
Ly$\alpha$ line of neutral hydrogen the central line absorption cross
section is
\begin{equation}
\chi_{{\rm Ly}\alpha}=2.0\times10^{-15}\;
       \left(\frac{300\;\mbox{km}\,\mbox{s}^{-1}}{v_D}\right)
       \;\mbox{cm}^2\;.
\end{equation}
Hence, the column density of neutral hydrogen for a halo with given
$\tau$ is
\begin{equation}
N_{\rm H}=0.5\times10^{14}\;\tau\;
        \left(\frac{v_D}{300\;\mbox{km}\,\mbox{s}^{-1}}\right)
        \;\mbox{cm}^{-2}.
\end{equation}
We would like to emphasize that this value is the column density of
the neutral halo alone. Considering the column density of the source
region, the total column density is about twice as high and is in the
range $N_{\rm H}\sim [10^{13},10^{16}]\;\mbox{cm}^{-2}$ for the
investigated configurations with $\tau=[0.1,100]$. 

In comparison to column densities deduced from Voigt profile fitting
procedures of Ly$\alpha$ profiles of high redshift radio galaxies
(van Ojik et al. \cite{vanojik:etal97}), our values are at least an
order of magnitude too low. With the present implementation of our
finite element code, we could calculate systems with column densities
up to $N_{\rm H}\sim10^{17}-10^{18}\;\mbox{cm}^{-2}$, but only with a
large amount of computational time. Still higher column densities are
beyond feasibility.

\begin{table}
\caption[]{Parameters used for all calculations. Distances are given
in units of the computational cube (see Sect.~\ref{normalization}).}
\begin{tabular}{cccccccc}
\hline\noalign{\smallskip}
$r_h$ & $r_c$ & $\alpha$ & $r_s$ & $v_D$ & $v_0$ & $r_0$ & $R_0$ \\
\noalign{\smallskip}
\hline\noalign{\smallskip}
1.0 & 0.2 & $10^3$ & 0.2 & $10^{-3} c$ & $-10^{-3} c$ & 0.2 & 1.0\\
\noalign{\smallskip}\hline
\end{tabular}
\label{parameters_table}
\end{table}

\begin{figure*}
\centering
\includegraphics*[width=17cm,bb=63 398 504 716]{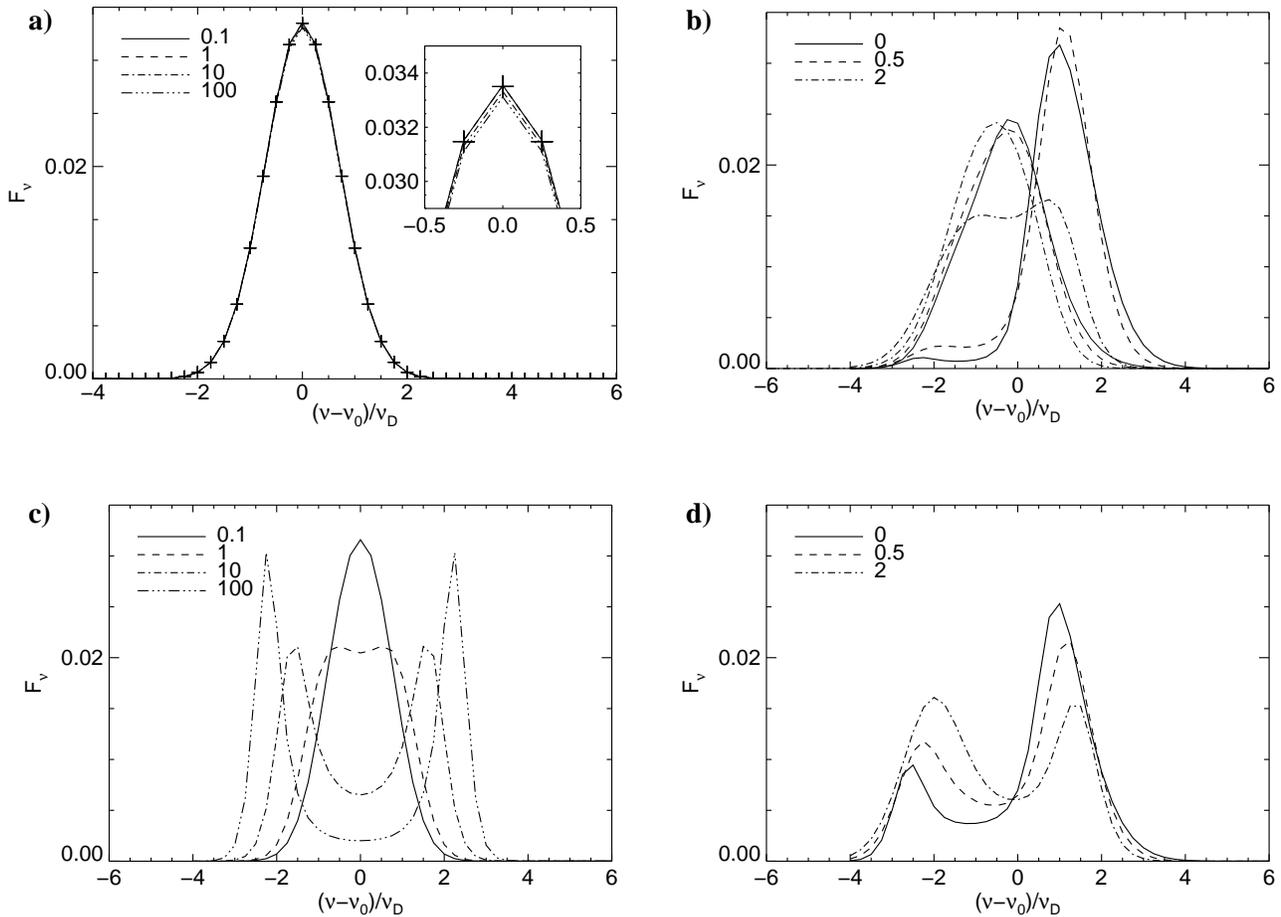}
\caption{Ly$\alpha$ line profiles calculated with the FE code for a
spherically symmetric model configuration: a) a static halo with
coherent scattering, b) an infalling halo with coherent scattering, c)
a static halo with complete redistribution, and d) an infalling halo
with complete redistribution. For the static cases a) and c) the line
styles refer to calculations with different optical depth $\tau$ as
indicated. The small window in a) enlarges the peak of the line. The
crosses mark the results of the analytical solution. For the moving
halos we show in b) the results for $\tau=1$ (thin lines) and
$\tau=10$ (thick lines) and in d) only for $\tau=10$. Here, the line
styles refer to the exponent $l$ used for the velocity fields.  }
\label{vergleich}
\end{figure*}

\begin{figure*}
\centering
\includegraphics*[width=17cm,bb=41 275 586 702]{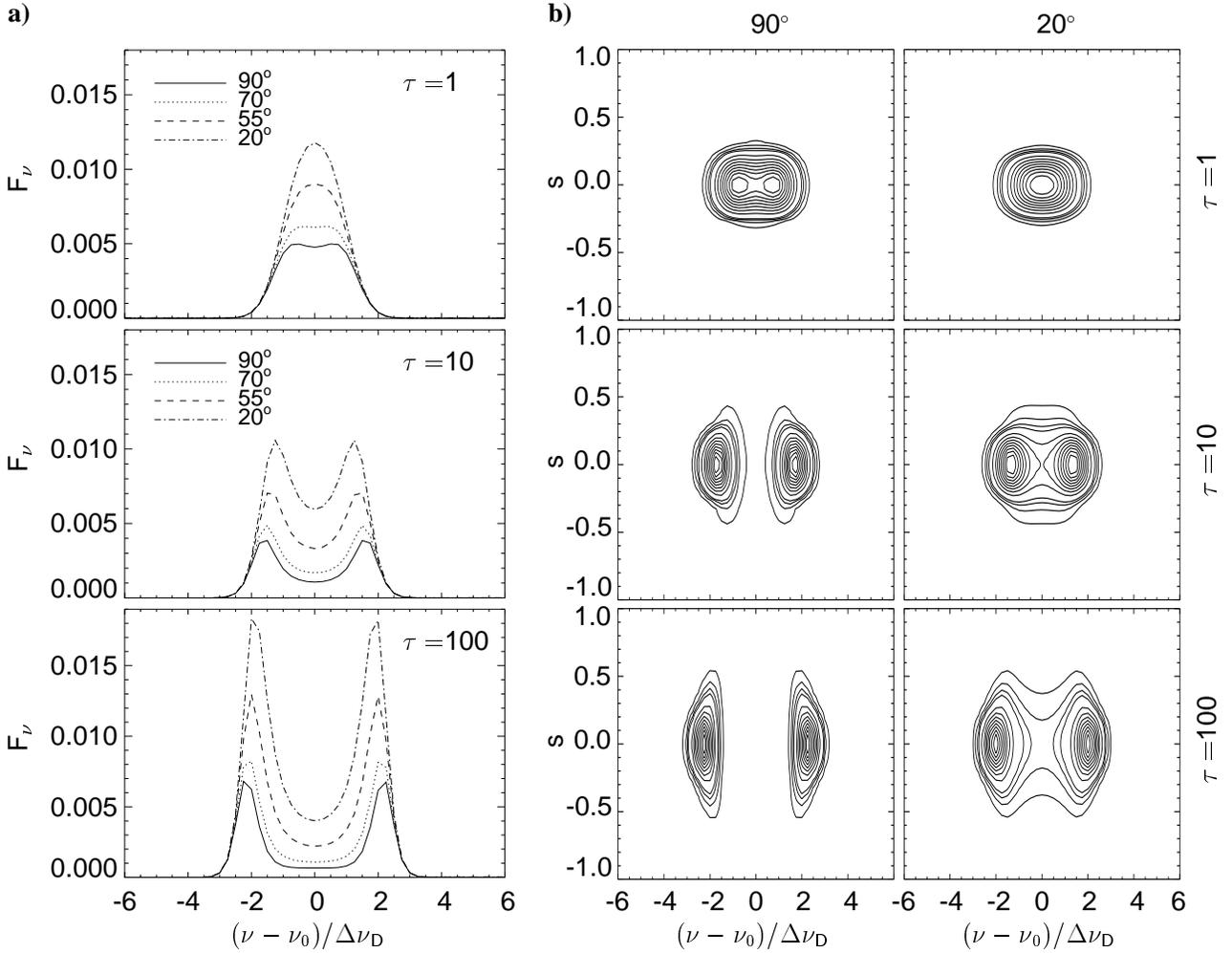}
\caption{Results obtained for a static, disk-like model configuration:
a) Emergent line profiles for different viewing angles and optical depths
$\tau=\tau(\n_{\rm thick})$. b) Two-dimensional spectra for 
an edge-on view (90$^\circ$) and a nearly face-on view (20$^\circ$)
and different optical depths. The position and width of the slit is
indicated in Fig.~\ref{srs_oblate}a. The contours are given for
2.5\%, 5\%, 7.5\%, 10\%, 20\%, ..., 90\% of the maximum value.}
\label{static_spectren_slit}
\end{figure*}

\begin{figure*}
\centering
\includegraphics*[width=17cm,bb=12 167 573 785]{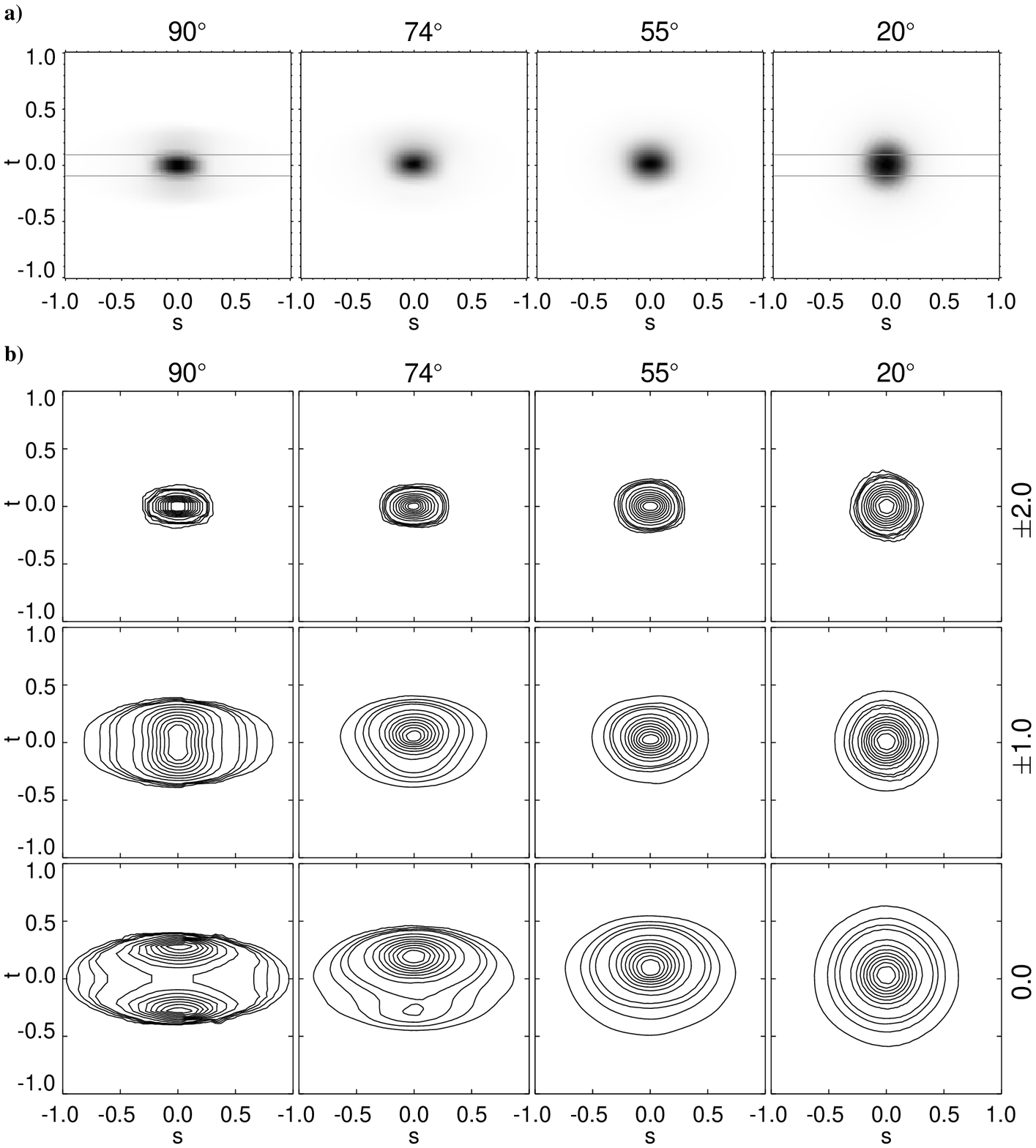}
\caption{Spatial intensity distribution for a static disk-like model
configuration with $\tau(\n_{\rm thick})=10$ for different viewing
angles: a) Frequency integrated intensity distribution. b) Intensity
distribution for specific frequencies. The frequency is given in
Doppler units relative to the line center $(\nu-\nu_0)/\Delta\nu_{\rm
D}$ at the right hand side.}
\label{srs_oblate}
\end{figure*}

\section{Test calculations}

We start with a spherically symmetric model configuration ($a=b=c=1$)
and a single source region located at $\x=0$. Fig.~\ref{vergleich}
shows the results of the finite element code (FE code) for different
optical depths, velocity fields and redistribution functions.  We used
41 frequencies equally spaced in the interval
$(\nu-\nu_0)/\Delta\nu_{D}=[-4,6]$ and 80 directions. Starting with a
grid of $4^3$ cells and a pre-refined source region, we needed 3--5
spatial refinement steps.

The simplest case is a static model with coherent isotropic
scattering. Fig.~\ref{vergleich}a displays the emergent line profiles
for different $\tau$. As expected, the Doppler profile is preserved
and the flux $F_{\nu}$ is independent on $\tau$. The deviation of the
numerical results from the analytical solution indicated with crosses
is very small. The line profiles for $\tau=0.1$ and $\tau=1$ are
identical even in the little window which shows the peak of the line
in more detail. For $\tau=100$, the total flux is still conserved
better than 99\%. This result demonstrates that the frequency-dependent
version of our FE code works correctly. 

Next, we consider an infalling halo with coherent scattering and show
the effects of frequency coupling due to the Doppler term.  The
emergent line profiles in Fig.~\ref{vergleich}b are plotted for
different exponents $l$ of the velocity field $\v_{\rm io}$ defined in
Eq.~(\ref{vio_eq}). The line profiles are redshifted for $\tau=1$
(thin lines). Most of the photons directly travel through the halo
moving away from the observer. Since the Doppler term is proportional
to the gradient of the velocity field, the redshift is larger for a
greater exponent $l$. For $\tau=10$ (thick lines), the line profiles
are blueshifted. Before photons escape from the optically thick halo
in front of the source, they are back-scattered and blueshifted in
the approaching halo behind the source. The blueshift is less
pronounced for the accelerated infall with $l=2$, because the strong
gradient of the velocity field leads to a slight redshift in the very
inner parts of the halo. In this region, the total optical depth is
still small. Further out, where the total optical depth increases, the
line profile becomes blueshifted.

Complete redistribution is another method of frequency coupling which
leads to a stronger coupling than the Doppler effect (see Sect.~3).
The line profiles obtained for a static model with complete
redistribution are displayed in Fig.~\ref{vergleich}c for different
$\tau$. With increasing optical depth the photons more and more escape
through the line wings. For $\tau\ge 1$, we get a double-peaked line
profile with an absorption trough in the line center. The greater
$\tau$ the larger the distance between the peaks and the depth of the
absorption trough. Since our frequency resolution is too poor for the
pointed wings, the flux conservation is only 96\% for $\tau=100$.

Fig.~\ref{vergleich}d shows the results for an infalling halo with
complete redistribution for $\tau=10$ and different exponents $l$.
For $l=0$ and $l=0.5$ the infalling motion of the halo enhances the
blue wing of the double-peaked line profile. Equally, an outflowing
halo would enhance the red peak. But for $l=2$, the red peak is
slightly enhanced for an infalling halo due to the strong velocity
gradient, as explained above. This example affords an insight into the
mechanisms of resonance line formation and shows the necessity of a
profound multi-dimensional treatment.

\section{Applications}

All calculations discussed in this section were performed with
complete redistribution. We used 49 equidistant frequencies covering
the interval $(\nu-\nu_0)/\Delta\nu_D=[-6,6]$, 80 directions and
started from a grid with $4^3$ cells and pre-refined source regions,
which results in several $10^3$ cells for the initial mesh.
3--7 refinement steps were performed leading to approximately
$10^5$ cells for the finest grid.

\subsection{Elliptical Halos}

\begin{figure*}
\centering
\includegraphics*[width=17cm,bb=41 275 586 702]{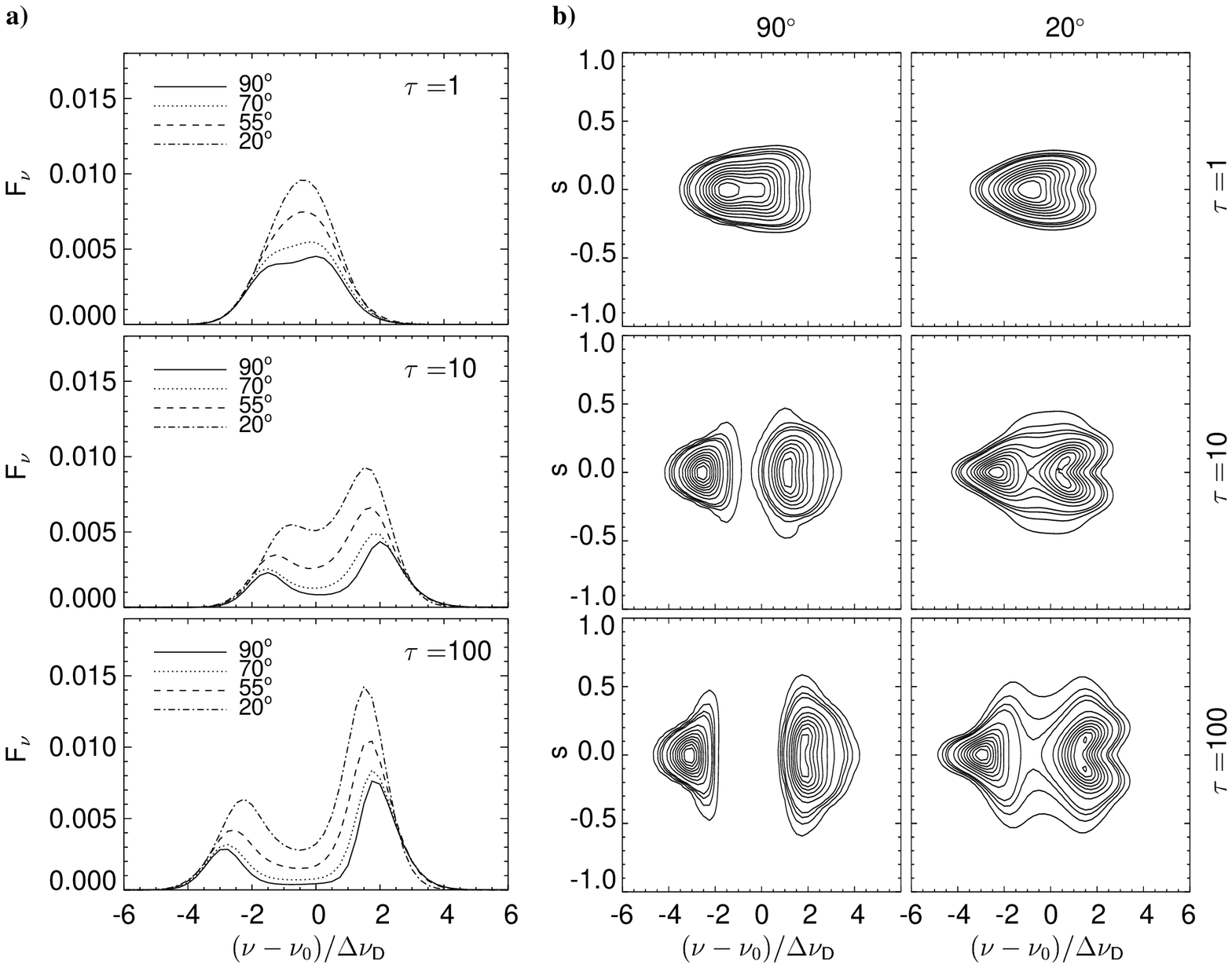}
\caption{Results obtained for a disk-like model configuration with
global infall motion ($l=0.5$): a) Emergent line profiles for
different viewing angles and optical depths $\tau=\tau(\n_{\rm
thick})$. b) Two-dimensional spectra for an edge-on view (90$^\circ$)
and a nearly face-on view (20$^\circ$) and different optical
depths. The position and width of the slit is indicated in
Fig.~\ref{srs_oblate}a. The contours are given for 2.5\%, 5\%, 7.5\%,
10\%, 20\%, ..., 90\% of the maximum value.}
\label{infall_spectren_slit}
\end{figure*}

\begin{figure*}
\centering
\includegraphics*[width=17cm,bb=41 399 586 702]{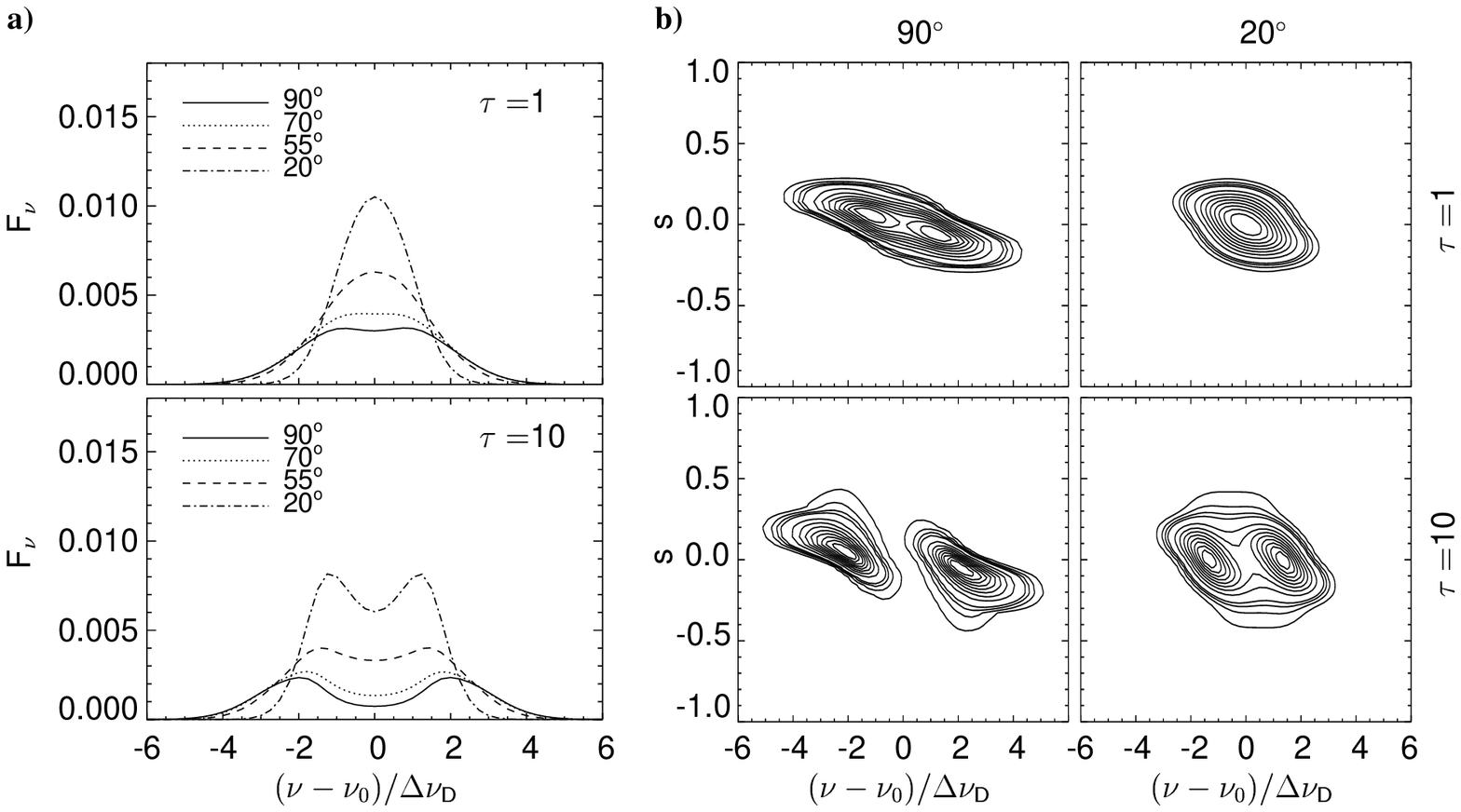}
\caption{Results obtained for a disk-like model configuration with
Keplerian rotation ($l=0.5$) around the $z$-axis: a) Emergent line
profiles for different viewing angles and optical depths
$\tau=\tau(\n_{\rm thick})$. b) Two-dimensional spectra for an edge-on
view (90$^\circ$) and a nearly face-on view (20$^\circ$) and different
optical depths. The position and width of the slit is indicated in
Fig.~\ref{srs_oblate}a. The contours are given for 2.5\%, 5\%, 7.5\%,
10\%, 20\%, ..., 90\% of the maximum value.}
\label{rot_spectren_slit}
\end{figure*}

\begin{figure*}
\centering
\includegraphics*[width=17cm,bb=12 167 573 785]{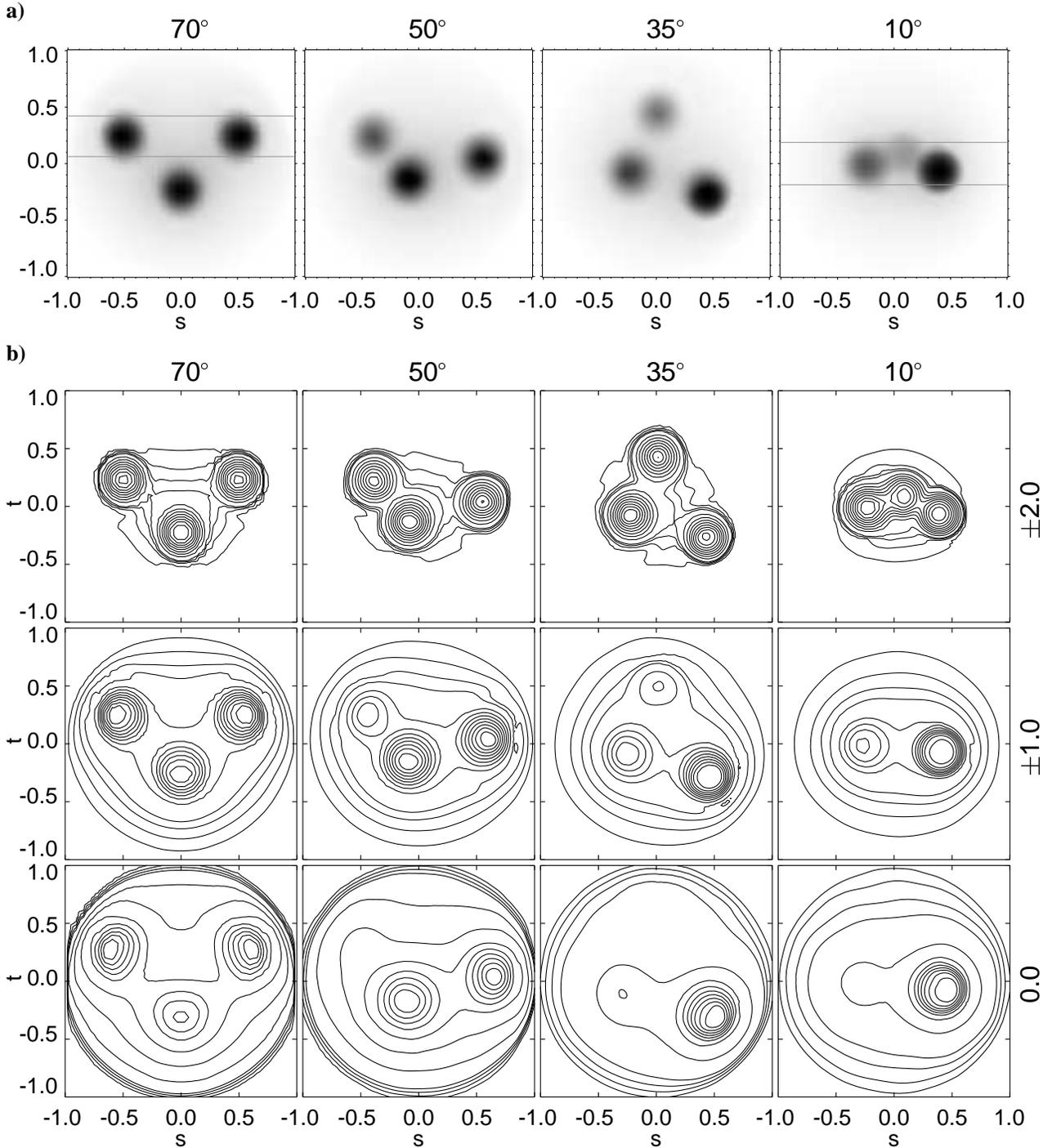}
\caption{Spatial intensity distribution for a static spherically
symmetric model configuration with $\tau=10$ and three source regions for
different viewing angles: a) Frequency integrated intensity
distribution. b) Intensity distribution for specific frequencies. The
frequency is given in Doppler units relative to the line center
$(\nu-\nu_0)/\Delta\nu_{\rm D}$ at the right hand side.}
\label{srs_three}
\end{figure*}

\begin{figure*}
\centering
\includegraphics*[width=17cm,bb=41 275 586 702]{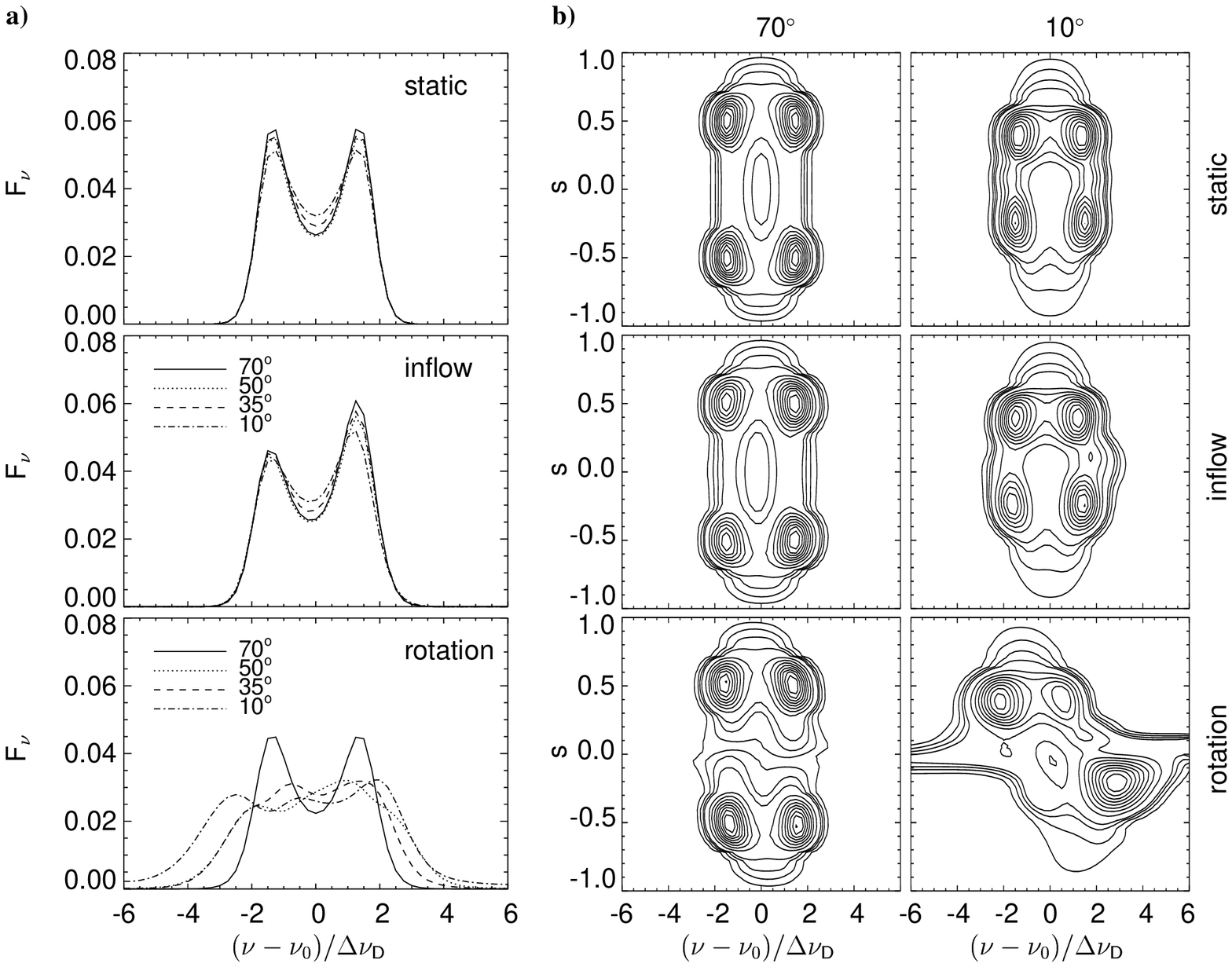}
\caption{Results obtained for a spherically symmetric model
configuration with $\tau=10$ and three source regions for the static
case and two different velocity fields (inflow and rotation): a)
Emergent line profiles for different viewing angles. b)
Two-dimensional spectra for a nearly face-on view (70$^\circ$) and a
nearly edge-on view (10$^\circ$). The position and width of the slit
is indicated in Fig.~\ref{srs_three}a. The contours are given for
2.5\%, 5\%, 7.5\%, 10\%, 20\%, ..., 90\% of the maximum value.}
\label{three_spectren_slit}
\end{figure*}

As a first step towards a full three-dimensional problem without any
symmetries, we investigated an axially symmetric, disk-like model
configuration ($a:b:c=3:3:1$) with a single source located at $\x=0$
for several $\tau=\tau(\n_{\rm thick})$ and for different velocity
fields. The most optical thick direction $\n_{\rm thick}$ is the
direction within the equatorial plane of the disk. The direction
perpendicular to the equatorial plane is the $z$-axis which we also
call rotational axis even for cases without rotation.

By means of the static case we show which kind of information can be
obtained from the results of our FE code. The emergent line profiles
provide global information on the underlying system. They are
displayed in Fig.~\ref{static_spectren_slit}a for different optical
depths and viewing angles. The viewing angle is defined as the angle
between the rotational axis and the direction towards the
observer. For instance, a viewing angle of 90$^\circ$ means, the disk
is ``observed'' edge-on. Again, we obtain the characteristical
double-peaked line profile for $\tau\ga 1$. The optical thickness
decreases for smaller viewing angles. For a viewing angle of $0^\circ$
the effective optical depth is only $\tau(\n_{\rm
thick})/3$. Consequently, the distance of the peaks in the line
profile and the depth of the absorption feature is growing with
increasing viewing angle. Note, that the total relative flux
$F(\n=0^{\circ})/F(\n=90^{\circ})$ escaping along the $z$-axis is
increasing with increasing $\tau(\n_{\rm
thick})$. Fig.~\ref{static_spectren_slit}b will be explained later.

Far more information is contained in the spatial distribution of the
line intensity projected into the $s$-$t$-plane which is the plane
perpendicular to the viewing direction. In Fig.~\ref{srs_oblate}a the
frequency-integrated intensity distribution is shown for $\tau=10$ and
different viewing angles. At 90$^{\circ}$ the Ly$\alpha$ image shows
an elliptical emission line region. The diffuse halo of scattered
radiation is most clearly visible. With decreasing viewing angle the
diffuse halo becomes less pronounced. For a nearly face-on view
(20$^{\circ}$), the halo is optically thinner and the intensity
distribution appears spherically symmetric.

Fig.~\ref{srs_oblate}b displays the spatial distribution of the
Ly$\alpha$ intensity for different viewing angles (columns) at
selected frequencies (rows). For the edge-on view, the intensity
distribution strongly depends on the frequency.  In the outer line
wings at $(\nu-\nu_0)/\Delta\nu_D=\pm2$ the contours are elliptical
and reproduce the disk-like form of the source region. The peaks of
the line profile are at $(\nu-\nu_0)/\Delta\nu_D\sim\pm1$. Here, the
major axis of the elliptical intensity distribution in the inner parts
of the model is parallel to the rotational axis, whereas the major
axis of the elliptical contour lines in the outer parts remains
parallel to the equatorial plane of the disk. In the very center of
the line ($(\nu-\nu_0)/\Delta\nu_D=0$), the direct view to the source
region is blocked by optically thick material. There appear two knots
of Ly$\alpha$ emission directly above and below the disk, which are
surrounded by an extended elliptical halo. The Ly$\alpha$ knot below
the disk vanishes with decreasing viewing angle.  At 20$^{\circ}$ the
intensity distribution is spherically symmetric for all
frequencies. The observable emission region is most extended at
the frequency of the line center.

Two-dimensional spectra from high-resolution spectroscopy provide
frequency-dependent data for only one spatial direction. To be able to
compare our results with these observations we calculated
two-dimensional spectra using the data within the slits indicated in
Fig.~\ref{srs_oblate}a for the edge-on and nearly face-on view. The
results are shown in Fig.~\ref{static_spectren_slit}b for different
optical depths. We find that the form of the two-dimensional spectra
is relatively insensible to the width of the slit.  For $\tau=1$ a
single emission region is visible. But already at 90$^{\circ}$, the
highest contour lines reproduce the faint absorption trough of the
corresponding line profile (Fig.~\ref{static_spectren_slit}a). The
higher the optical depth the wider the gap and the spatial extent of
the two peaks. Note that the spatial extent of the outer contour line
only depends on $\tau(\n_{\rm thick})$ and not on the viewing
direction.

What changes when imposing a macroscopic velocity field is shown in
the following two figures. First, we consider an infalling velocity
field with $l=0.5$ suitable for a gravitational collapse.
Fig.~\ref{infall_spectren_slit}a displays the calculated line profiles
for different $\tau$ and viewing angles. As expected, the blue peak of
the line is enhanced. The higher $\tau$ the stronger the blue
peak. Fig.~\ref{infall_spectren_slit}b shows the corresponding
two-dimensional spectra obtained with the same slit width and position
as in the static case. For low optical depth, the shape of the contour
lines is a triangle. Photons changing frequency in order to escape via
the blue wing are also scattered in space. The consequence is the
greater spatial extent of the blue wing. Apart from a growing gap
between the two peaks with higher optical depth, the general
triangular shape in conserved.

Next, we investigated the elliptical model configuration with
Keplerian rotation ($l=0.5$), where the $z$-axis is the axis of
rotation. The results are plotted in Fig.~\ref{rot_spectren_slit} for
$\tau=1$ and $\tau=10$. The emergent line profiles are symmetric with
respect to the line center and show the same behavior with growing
optical depth as in the static case. However, the extension of the
line wings towards higher and lower frequencies is strongly increasing
with increasing viewing angle because of the growing effect of the
velocity field. Rotation is clearly visible in the two-dimensional
spectra (Fig.~\ref{rot_spectren_slit}b). The shear of the contour
lines is the characteristical pattern indicating rotational
motion. For an edge-on view at $\tau=10$, Keplerian rotation produces
two banana-shaped emission regions.

In spite of the relatively low optical depth of the simple model
configurations, our results reflect the form of line profiles and the
patterns in two-dimensional spectra of many high redshift
galaxies. For example, the two-dimensional spectra of the Ly$\alpha$
blobs discovered by Steidel et al. (\cite{steidel:etal2000}, Fig.~8)
are comparable to the results obtained for infalling
(Fig.~\ref{infall_spectren_slit}b) and rotating
(Fig.~\ref{rot_spectren_slit}b) halos. In the sample of van Ojik et
al. (\cite{vanojik:etal97}) are many high redshift radio galaxies with
single-peaked and double-peaked Ly$\alpha$ profiles. The corresponding
two-dimensional spectra show asymmetrical emission regions which are
more or less extended in space. De Breuck et
al.~(\cite{debreuck:etal2000}) find in their statistical study of
emission lines from high redshift radio galaxies that the triangular
shape of the Ly$\alpha$ emission is a characteristical pattern in the
two-dimensional spectra of high redshift radio galaxies. Since the
emission of the blue peak of the line profile is predominately less
pronounced, most of the associated halos should be in the state of
expansion.

\subsection{Multiple Sources}   

High redshift radio galaxies are found in the center of proto
clusters. In such an environment, it could be possible that the
Ly$\alpha$ emission of several galaxies is scattered in a common
halo. To investigate this scenario, we started with a spherically
symmetric distribution for the extinction coefficient and with three
source regions located at $x_1=[0.5,0.25,0]$, $x_2=[-0.5,0.25,0]$ and
$x_3=[0,-0.25,0]$ forming a triangle in the $x$-$y$-plane. In the
following figures, the results for $\tau=10$ are presented.

We begin with the static model. Figure~\ref{srs_three}a shows
Ly$\alpha$ images for four selected viewing directions. The specified
angle is the angle between the viewing direction and the
$x$-$y$-plane. Note that the orientation of the source positions
within the plane is different for each image. Viewing the
configuration almost perpendicular to the $x$-$y$-plane
(70$^{\circ}$), renders all three source regions visible, because the
source regions are situated in the more optically thin, outer parts
of the halo. Remember that most of the scattering matter is in and
around the center of the system. For other angles, some of the source
regions are located behind the center and therefore less visible on
the images. Figure~\ref{srs_three}b shows images at different
frequencies. In the line wings at $(\nu-\nu_0)/\Delta\nu_D=\pm2$, the
number and position of the source regions can be determined for all
viewing angles. However, at frequencies around the line center at
$(\nu-\nu_0)/\Delta\nu_D=\pm1\;\mbox{and}\;0$, the number of visible
sources depends on the viewing angle. Some of the images show only one,
other two or three sources.

The corresponding line profiles and two-dimensional spectra are
displayed in Fig.~\ref{three_spectren_slit} for the static case as
well as for a halo with global inflow and Keplerian rotation. Width
and position of the slits are depicted in Fig.~\ref{srs_three}a. We
get double-peaked line profiles for almost all cases, except for the
rotating halo, where the line profiles are very broad for viewing
angles lower than 70$^{\circ}$. Additional features, dips or
shoulders, are visible in the red or blue wing. They arise because the
three sources have significantly different velocities components with
respect to the observer.

The slit for a viewing angle of 70$^{\circ}$ contains two
sources. They show up as four emission regions in the two-dimensional
spectra (Fig.~\ref{three_spectren_slit}b). The pattern is very
symmetric, even for the moving halos. For a viewing angle of
10$^{\circ}$, the slit covers all source regions. Nevertheless, the
two-dimensional spectra are dominated by two pairs of emission regions
resulting from the sources located closer to the observer. The third
source region only shows up as a faint emission in the blue part in the
case of global inflow. In the case of rotation, emission regions from
a third source are present. But the overall pattern is very irregular
and prevents a clear identification of the emission regions.

This example demonstrates the complexity of three dimensional
problems. In a clumpy medium, the determination of the number and
position of Ly$\alpha$ sources would be difficult by means of
frequency integrated images alone. Two-dimensional spectra may help,
but could prove to be too complicated. More promising are images
obtained from different parts of the line profile or information from
other emission lines of OIII or H$\alpha$ in a manner proposed by
Kurk et al.~(\cite{kurk:etal2001}).

\section{Summary}

We presented a finite element code for solving the resonance line
transfer problem in moving media. Non-relativistic velocity fields and
complete redistribution are considered. The code is applicable to any
three-dimensional model configuration with optical depths up to
$10^{3-4}$.

We showed applications to the hydrogen Ly$\alpha$ line of slightly
optically thick model configurations ($\tau\le 10^2$) and discussed
the resulting line profiles, Ly$\alpha$ images and two-dimensional
spectra. The systematic approach from very simple to more complex
models gave the following results:
\begin{itemize}
\item An optical depth of $\tau\ga 1$ leads to the characteristic
double peaked line profile with a central absorption trough as
expected from analytical studies (e.g. Neufeld \cite{neufeld90}). This
form of the profile is the result of scattering in space and
frequency. Photons escape via the line wings where the optical depth
is much lower.
\item Global velocity fields destroy the symmetry of the line
profile. Generally, the blue peak of the profile is enhanced for
models with infall motion and the red peak for models with outflow
motion. But there are certain velocity fields (e.g.~with steep
gradients) and spatial distributions of the extinction coefficient,
where the formation of a prominent peak is suppressed.
\item Double-peaked line-profiles show up as two emission regions in
the two-dimensional spectra. Global infall or outflow leads to an
overall triangular shape of the emission. Rotation produces a shear
pattern resulting in banana-shaped emission regions for optical
depths $\ga 10$.
\item For non-symmetrical model configurations, the optical depth
varies with the line of sight. Thus, the total flux, the depth of the
absorption trough and the pattern in the two-dimensional spectra
strongly depend on the viewing direction.
\end{itemize}

The applications demonstrate the capacity of the finite element code
and show that the three-dimensional structure and the kinematics of
the model configurations are very important. Thus, beside exploring
higher optical depths up to the limits of our code, we will consider
clumpy density distributions as well as dust absorption and try to
model the Ly$\alpha$ emission of individual high redshift galaxies. In
addition, we intend to implement a second order method for the
frequency discretization. Furthermore, we plan to overcome the
difficulties in solving the line transfer problem for configurations
with $\tau>10^4$.  Therefore, we will extend our method and use the
diffusion approximation in the most optically thick regions. In the
other regions, the full frequency-dependent line transfer problem must
be solved.

\begin{acknowledgements}
We would like to thank Rainer Wehrse and Guido Kanschat for useful
discussions. This work is supported by the Deutsche
Forschungsgemeinschaft (DFG) within the SFB 359 ``Reactive Flows,
Diffusion and Transport'' and SFB 439 ``Galaxies in the young
Universe''.
\end{acknowledgements}


\begin{thebibliography}{}

     \bibitem[1972]{adams72}
      Adams, T.F.\ 1972, ApJ 174, 439

     \bibitem[2001]{ahn:etal2001}
      Ahn, S.-H., Lee, H.-W., \& Lee, H. M.\ 2001, ApJ 554, 604

     \bibitem[2002]{ahn:etal2002}
      Ahn, S.-H., Lee, H.-W., \& Lee, H. M.\ 2002, ApJ 567, 992

     \bibitem[1999]{baschek:wehrse99}
      Baschek, B., \& Wehrse, R.\ 1999, Physics Reports 311, 201

     \bibitem[2000]{bicknell:etal2000}
      Bicknell, G.V., Sutherland, R.S., van Breugel, W.J.M., et al.\ 2000
      ApJ 540, 678

     \bibitem[2000]{debreuck:etal2000}
      De Breuck, C., R\"ottgering, H., Miley, G., van Breugel, W.,
      \& Best, P.\ 2000, A\&A 362, 519

     \bibitem[2000]{fynbo:etal2000}
      Fynbo, J.U., Thomsen, B., \& M{{\o}}ller, P.\ 2000, A\&A 353, 457

     \bibitem[2001]{fynbo:etal2001}
      Fynbo, J.U., M{{\o}}ller, P., \& Thomsen, B.\ 2001, A\&A 374, 443
     
     \bibitem[1993]{hippelein:meisenheimer93}
      Hippelein, H., \& Meisenheimer, K.\ 1993, Nature 362, 224

     \bibitem[1998]{hu:etal98}
      Hu, E.M., Cowie, L.L., \& McMahon, R.G.\ 1998, ApJ 502, L99

     \bibitem[1980]{hummer:kunasz80}
      Hummer, D.G., \& Kunasz, P.B.\ 1980, ApJ 236, 609

     \bibitem[1996]{kanschat96} 
      Kanschat, G.\ 1996, Ph.D.\ Thesis, Univ.\ of Heidelberg

     \bibitem[2001]{kurk:etal2001}
      Kurk, J.D., R\"ottgering, H.J.A., Miley, G.K.,
      \& Pentericci, L.\ 2001, RevMexAA (Serie de Conferencias), in press

     \bibitem[2000]{kudritzki:etal2000}
      Kudritzki, R.-P., Mendez, R.H., Feldmeier, J.J., et al.\ 
      2000, ApJ 536, 19

     \bibitem[1984]{mihalas:weibel-mihalas84}
      Mihalas, D., \& Weibel-Mihalas, B.\ 1984, Foundation of Radiation
      Hydrodynamics, Oxford University Press, New York

     \bibitem[1990]{neufeld90}
      Neufeld, D.A.\ 1990, ApJ 350,216


     \bibitem[2001]{richling:etal2001}
      Richling, S., Meink\"ohn, E., Kryzhevoi, N., \& Kanschat, G.\ 2001,
      A\&A 380, 776 (Paper I)

     \bibitem[2000]{rhoads:etal2000}
      Rhoads, J.E., Malhotra, S., \& Dey, A.\ 2000, ApJ 545, L85

     \bibitem[2000]{steidel:etal2000}
      Steidel, C.C., Adelberger, K.L., Sharply, A.E., et al.\ 
      2000, ApJ 532, 170

     \bibitem[1996]{vanojik:etal96}
      van Ojik, R., R\"ottgering, H.J.A., Carilli, C.L., et al.\
      1996, A\&A 313, 25

     \bibitem[1997]{vanojik:etal97}
      van Ojik, R., R\"ottgering, H.J.A., Miley, G.K., \& Hunstead, R.W.\ 1997,
      A\&A 317, 358

     \bibitem[1999]{villar-martin:etal99}
      Villar-Mart{\'{\i}}n, M., Binette, L., \& Fosbury, R.A.E.\ 1999,
      A\&A 346, 7

     \bibitem[2000]{wehrse:etal2000}
      Wehrse R., Baschek B., \& von Waldenfels, W.\ 2000, A\&A 359, 780

  \end{thebibliography}
\end{document}